\documentclass[fleqn,usenatbib]{mnras}

\usepackage[T1]{fontenc}
\usepackage{ae,aecompl}
\usepackage{bm}
\usepackage{graphicx}  
\usepackage{verbatim}
\usepackage{cprotect}

\usepackage{amsmath}	
\usepackage{amssymb}	
\usepackage{txfonts}

\title[Minimising numerical viscosity in SPH simulations]{Minimising the numerical viscosity in Smoothed Particle Hydrodynamics simulations of discs}

\author[Chen \& Nixon]{Cheng Chen\thanks{Email: C.Chen6@leeds.ac.uk} and C.~J. Nixon
\\ School of Physics and Astronomy,  University of Leeds, Sir William Henry Bragg Building, Woodhouse Ln., Leeds LS2 9JT, UK\\}

\date{Accepted 2025 April 28. Received 2025 April 24; in original form 2024 September 04}

\pubyear{2024}

\begin{document}
\label{firstpage}
\pagerange{\pageref{firstpage}--\pageref{lastpage}} 
\maketitle


\begin{abstract}
Simulations using the Smoothed Particle Hydrodynamics (SPH) technique typically include numerical viscosity to model shocks and maintain particle order on the kernel scale. This numerical viscosity is composed of linear and quadratic terms, with coefficients $\alpha_{\rm SPH}$ and $\beta_{\rm SPH}$ respectively. Setting these coefficients too high results in excessive numerical dissipation, whereas setting them too low may lead to unwanted effects such as particle penetration, which also leads to excess dissipation.  In this study, we simulate accretion discs using the SPH code {\sc phantom} to investigate the effective disc viscosity arising from numerical viscosity. We model steady-state coplanar and circular discs with different values of $\alpha_{\rm SPH}$ and $\beta_{\rm SPH}$, from which we determine the coefficients that lead to minimum levels of numerical viscosity by maximising the steady-state disc surface density for the same mass input rate. We find that, for planar and circular discs, the default values of the numerical viscosity parameters in the {\sc phantom} code can be too high particularly for the quadratic term. As higher values of the coefficients are required to adequately capture strong shocks in the flow, we suggest that the coefficient of the quadratic term should be time-dependent in a similar manner to the presently used ``switches'' on the linear term. This can be simply achieved by setting $\beta_{\rm SPH}$ to be a constant multiple of $\alpha_{\rm SPH}$ with $\alpha_{\rm SPH}$ determined by an appropriate switch, as previously advocated in the literature.
\end{abstract}

\begin{keywords}
accretion, accretion discs --- methods: numerical --- hydrodynamics --- shock waves
\end{keywords}

\section{Introduction}
Astrophysical flows come in many flavours, from small scale phenomena 
\citep[e.g. convective granulation on the surface of stars; ][]{Dravins1987, Mathur2011, Beeck2013} to large scale processes (e.g. bubbles heating the medium surrounding galaxy clusters; e.g. \citealt{Ruszkowski2002}). In many cases astrophysical flows take the form of an accretion disc, which comprises a centrifugally supported flow orbiting a central object \citep[e.g.][]{Pringle1981}. These discs occur in star forming regions around young stars, in stellar binary systems including discs formed by the transfer of mass from one star to another, and around the supermassive black holes in the centres of galaxies. In the simplest case the disc possesses a high degree of symmetry, with smooth, circular and planar orbits, which makes analytical progress easier. However, in nature it is common for discs to be eccentric and/or warped, making analytical progress difficult \citep[see, e.g.,][]{Ogilvie2006}. It is therefore common to resort to numerical hydrodynamical simulations to make progress in understanding the dynamics of accretion discs (and indeed astrophysical flows in general).

There are now several distinct methods for numerical hydrodynamical simulations that are routinely employed in astrophysics. Eulerian (or grid) based methods, which discretise the computational domain and compute the flow properties through regularly spaced cells, are perhaps the most developed methods. Lagrangian (or particle) based methods, which discretise the fluid by mass, are regularly employed in cases where the flow is geometrically complex. More recently hybrid methods \citep[e.g.][]{Springel2011, Hopkins2015} have gained popularity. Here we are focussed on the Smoothed Particle Hydrodynamics technique of \cite{Lucy1977} and \citet[][see \citealt{Price2012a} for a review]{Gingold1977}, and in particular the version of SPH implemented in the {\sc phantom} SPH code \citep{Price2018}. {\sc phantom} has been used to model a variety of astrophysical flows from turbulence in star forming regions \citep[e.g.][]{Price2010} to jets from collapsing stars \citep[e.g.][]{Price2012c} to stars being disrupted by supermassive black holes \citep[e.g.][]{Norman2021}. However, the main use of {\sc phantom} has been to model the dynamics of accretion discs, starting with \citep{Lodato2010} who used the code to model warped discs. 

Numerical simulations of fluids typically include either an explicit artificial viscosity or employ a Riemann solver to capture the effects of discontinuities such as shocks. Both methods result in numerical viscosity. In SPH simulations, an explicit artificial viscosity is the standard approach \citep{Monaghan1992}, although different types of SPH algorithms have employed a Rieman solver (or Godonov) approach as well \citep[e.g.][]{Inutsuka2002,Hopkins2015}. The artificial viscosity introduced to the SPH algorithm is composed of two terms; (1) a term that is linear in the velocity difference between particles and has a coefficient $\alpha_{\rm SPH}$, and (2) a term that is quadratic in the velocity difference with coefficient $\beta_{\rm SPH}$. For a useful pedagogical discussion of these terms see \cite{Cossins2010}. Both of these terms contribute \citep[e.g.][]{Cullen2010} to the smooth running of an SPH code, principally by helping to maintain good particle order (required to ensure the particles provide an accurate representation of the fluid and particularly needed in regions of strong shear and/or shocks) and by preventing overshooting and penetration of converging flows (with the quadratic term in particular needed for high Mach number shocks). However, both terms also introduce a (numerical, and thus unwanted) shear and bulk viscosity to the flow \citep[see][for discussion specifically related to discs]{Artymowicz1994,Murray1996,Lodato2010}. These effective viscosities arising from the numerical terms contribute to both angular momentum transport and heating in discs \citep[see, also,][]{Meru2012}, and in some cases the numerical effect can be larger than the corresponding physical effect that is being simulated. It is therefore important to minimise these terms as much as possible, while ensuring that they perform their helpful tasks as needed.\footnote{As the linear and quadratic viscosity terms are linearly and quadratically proportional to the resolution lengthscale, it is clear that once high enough resolution is reached the artificial viscosity can be reduced to required levels. However, as computational power is finite, it is not always possible to reach this level.}

In this work, we explore the values of the coefficients $\alpha_{\rm SPH}$ and $\beta_{\rm SPH}$ that minimise the effective viscosity arising from the artificial viscosity terms for accretion discs. We explore the simplest case of a planar, circular disc as this is the most straightforward for interpreting the results. We exploit the steady nature of the numerical solutions to infer the relative magnitude of the effective viscosity arising from the artificial viscosity terms for various values of the coefficients $\alpha_{\rm SPH}$ and $\beta_{\rm SPH}$. 

The structure of the paper is as follows. In Section~\ref{art} we describe the artificial viscosity terms in SPH. In Section~\ref{discs} we provide the background on accretion discs required to interpret the simulations. In Section~\ref{sims} we present the numerical simulations of steady state discs for different viscosity coefficients. Finally, we discuss our results and present conclusions in Sections~\ref{discussion} \& \ref{conclusion} respectively.

\section{Artificial Viscosity in SPH}
\label{art}

Artificial viscosity in SPH is typically added via explicit terms in the momentum equation \citep[e.g.][]{Monaghan1983, Monaghan1992}. The exact form of the terms and the adopted values of the viscosity coefficients have varied over the years, often due to the competing needs of minimising unwanted dissipation away from shocks and resolving strong shocks that can occur in astrophysical flows. Two notable updates on the initial formulation are: (1) a re-formulation of the terms in analogy with the Riemann solvers that are used in some grid-based codes \citep{Monaghan1997,Chow1997,Price2004}, and (2) the implementation of viscosity switches which evolve a time-dependent equation for the coefficients to reduce their magnitude away from shocks and increase it where shocks are detected \citep{Morris1997,Cullen2010}.\footnote{There is also the \cite{Balsara1995} viscosity switch which multiplies the artificial viscosity terms by a factor that is of order unity for compression-dominated flows and small for shear-dominated flows. A modified version of the \cite{Balsara1995} switch is included in the formulation presented in {\sc phantom} \citep{Price2018} as part of the \cite{Cullen2010} viscosity switch.}

For the simulations we present in the next section we employ the SPH code {\sc phantom}, so we provide the relevant equations for the numerical viscosity implemented in {\sc phantom} here \citep[see][for details]{Price2018}. In {\sc phantom}, the dissipation term is given by
\begin{equation}
\label{alphabeta}
    \Pi^{a}_{\rm shock} \equiv -\sum\limits_{b} m_{b}\left[ \frac{q^{a}_{ab}}{\rho^2_{a}\Omega_{a}} \nabla_{a} W_{ab}  \left(h_{a} \right)  + \frac{q^{b}_{ab}}{\rho^2_{b}\Omega_{b}} \nabla_{a} W_{ab} \left(h_{b} \right)  \right],
\end{equation}
where $m_{b}$ is the particle mass, $h$ is the smoothing length, $\rho$ is the density, $\Omega$ is a factor related to the variable nature of the smoothing lengths, and $W_{ab}$ is the smoothing kernel and the quantity $q_{ab}^a$ is given by
\begin{equation}
q^{a}_{ab} =\left\{ 
\begin{aligned}
 &-\frac{1}{2}\rho_{a}{\rm v}_{\rm sig,a}{\bm v_{ab}} \cdot {\bm \hat{r}_{ab}} \ & {\rm for} \ {\bm v_{ab}} \cdot {\bm \hat{r}_{ab}} \ < \ 0, \\
 &~~~~~0 &\ {\rm otherwise},\\ 
\end{aligned}
\right.
\end{equation}
where ${\bm v_{ab}} = {\bm v_{a}}-{\bm v_{b}}$, ${\bm \hat{r}_{ab}} = \left( {\bm r_{a}}- {\bm r_{b}} \right)/|{\bm r_{a}}- {\bm r_{b}}| $ is the unit
vector along the line of sight between the particles. The signal speed, ${\rm v}_{\rm sig}$, is given by
\begin{equation}
    {\rm v}_{{\rm sig,}a} = \alpha_{{\rm SPH},a}\,c_{{\rm s,}a} + \beta_{\rm SPH}\,|{\bm v_{ab}} \cdot {\bm \hat{r}_{ab}}|,
\end{equation}
where $c_{\rm s}$ is the sound speed. In this manner the quantity $q_{ab}^a$ is composed of two terms: (1) the linear term which has coefficient $\alpha_{\rm SPH}$ and is linearly proportional to the sound speed and the SPH equivalent of the divergence, ${\bm v}_{ab} \cdot {\bm r}_{ab}$, and (2) the quadratic term which has coefficient $\beta_{\rm SPH}$ and is proportional to the divergence squared (with no dependence on the sound speed).

Initial SPH simulations employed fixed coefficients for the artificial viscosity terms, with values of $\alpha_{\rm SPH}$ = 1 and $\beta_{\rm SPH}$ = 2 found to provide the best results \citep{Monaghan1989, Monaghan1992}. In contrast, motivated by minimising the numerical viscosity and the resulting entropy generation for SPH simulations of self-gravitating discs, \citet{Lodato2004, Rice2005, Meru2011b, Meru2012} employed $\alpha_{\rm SPH}$ = 0.1 and $\beta_{\rm SPH}$ = 0.2. \cite{Price2010} suggest that $\beta = 4$ is necessary to prevent particle penetration occurring for shocks with Mach number $\gtrsim 10$.\footnote{Although it is worth noting that the implementation in {\sc phantom} at the time, used by \cite{Price2010}, made use of a formalism that multiplied the $\beta_{\rm SPH}$ term by the value of $\alpha_{\rm SPH}$, and the authors employed the Morris-Monaghan switch for $\alpha_{\rm SPH}$ with a minimum value of 0.05 and a maximum value of unity. Thus the need for an increased value of $\beta_{\rm SPH}$ may have been due to the multiplication by $\alpha_{\rm SPH} \le 1$.} Similarly, some implementations employ a relationship between $\alpha_{\rm SPH}$ and $\beta_{\rm SPH}$, with typically $\beta_{\rm SPH} = 2\alpha_{\rm SPH}$ \citep[e.g.][]{Monaghan1997,Morris1997,Price2004,Price2010,Price2012a}, while others assign values separately \citep[e.g.][]{Meru2012,Price2018}.

In {\sc phantom}, the current default implementation for the numerical viscosity\footnote{Note that here we are not referring to the ``artificial viscosity for a disc'' method that is the default in the disc setup, which attempts to scale the numerical viscosity to a \cite{SS1973} viscosity based on the initial disc conditions.} is a linear term where the value of $\alpha_{\rm SPH}$ for each particle is set using a modified version of the time-dependent Cullen-Dehnen switch with a minimum value of zero and a maximum value of unity, and a quadratic term where the value of $\beta_{\rm SPH}=2$ is a constant (i.e. not multiplied by the variable $\alpha_{\rm SPH}$). See \cite{Price2018} for details.

Finally, while it is clear that the artificial viscosity terms generate some level of shear viscosity in the flow, it is worth noting that this is not the only source of numerical viscosity present. In shearing flows the particles can become disordered, and one of the key features of SPH is its ability to re-order the particles so that they maintain a good representation of the underlying flow \citep[e.g.][]{Monaghan2005}. This reordering can contribute to the numerical viscosity present in the simulation. This was demonstrated by \cite{Dehnen2012} who showed that, even with the Cullen-Dehnen viscosity switch, there was some viscosity generated by the reordering of particles in the Gresho-Chan vortex test. We can therefore expect that numerical viscosity is generated by maintaining particle order in SPH simulations of Keplerian shear flows. As these effects are generated on the kernel scale, we would expect them to be reduced as the resolution of the simulation is increased.

\section{Accretion discs}
\label{discs}

An accretion disc \citep{Pringle1981, Frank2002} in its simplest form is composed of material on circular obits around a central Newtonian point mass with the angular momentum vectors of each ring of the disc parallel to each other, i.e. the disc is planar, and with the disc mass sufficiently small that the effects of gas self-gravity can be safely neglected. The orbital velocity varies with  (cylindrical) radius, $R$, as ${\rm v}_\phi = \sqrt{GM/R}$, where $G$ is the gravitational constant and $M$ is the mass of the central star. The time taken to complete an orbit is $t_{\rm orb} = 2\pi/\Omega$ where the orbital frequency $\Omega = {\rm v}_\phi/R$. The vertical density profile of the disc, $\rho(z)$, is set by the competition of gravity from the central object and gas pressure. For an isothermal disc with pressure proportional to the density, the vertical profile is a Gaussian with a scale-height given by $H = c_{\rm s}/\Omega$, with $c_{\rm s}$ the local sound speed. From this the disc angular semi-thickness is $H/R = c_{\rm s}/{\rm v}_\phi$, and for thin discs with $H/R \ll 1$ the disc evolution in the radial and vertical directions can be separated, and the vertical structure taken to be hydrostatic.

The radial evolution of the disc is generated by the action of a viscosity, typically taken to be of the form $\nu = \alpha c_{\rm s} H$, where $\nu$ is the kinematic viscosity, and $\alpha$ is a dimensionless parameter \citep{SS1973}. The viscosity is thought to arise due to turbulence in the disc, and in low-mass discs that are sufficiently ionised this can be driven by the magneto-rotational instability \citep{Balbus1991}. Observations of the viscous decay of time-dependent discs, principally in the Dwarf Nova subtype of the Cataclysmic Variable stars, find that (when the disc is fully-ionised) $\alpha \sim 0.2-0.4$ \citep{King2007,Kotko2012,Martin2019}. In low-ionisation states the value of $\alpha$ may be much lower \citep[e.g.][]{Gammie1996,Gammie1998}. In sufficiently massive discs, turbulence can develop due to self-gravity that provides $\alpha \approx 0.1$ \citep[e.g.][]{Lodato2004}.

Together with the equations for conservation of mass and angular momentum, the viscosity results in a diffusion equation for the disc surface density $\Sigma(R,t)$ \citep[e.g.][]{Pringle1981}. Depending on the location of the inner and outer boundaries to the disc, the boundary conditions applied there,\footnote{Typically the boundary conditions are taken to be zero viscous torque, such that the surface density goes to zero at the boundary. In this case mass can flow freely across the boundary carrying with it its angular momentum. A non-zero torque boundary may be more appropriate in some discs due to a variety of different physical effects \citep[see][for discussion]{Nixon2021}.} and the assumed viscosity law, it is possible to derive analytical solutions for the time-dependent evolution of the disc \citep[e.g.][]{LP1974,Tanaka2011,Nixon2021}. 

Previous works have measured the numerical viscosity by comparing the viscous spreading of a ring of particles in a simulation to that expected from the Green's function solution of \citet[][e.g. \citealt{Murray1996}, \citealt{Lodato2004}]{LP1974}. For these tests it is customary to turn off the pressure force (to avoid the ring spreading due to the large pressure gradients at the edges), and as the ring spreads the resolution (determined by the local particle number density) will not remain constant. We wish to explore the level of numerical viscosity present in a controlled environment that closely resembles that employed in typical disc simulations with SPH. For this we employ steady-state discs.

To produce a steady state disc, we exploit the mass injection into SPH discs described, and implemented into {\sc phantom}, by \cite{Drewes2021}. Using this method mass leaves the simulation domain through inner and outer boundaries at radii denoted $R_{\rm in}$ and $R_{\rm out}$ respectively, and mass is replenished by the addition of particles to the disc. Hence, by evolving the simulation for a sufficiently long timescale, the disc can achieve a steady state, where the mass reaching the inner and outer disc boundaries is offset by the input of mass at the injection radius.

For such steady state discs, and using the 1D diffusion equation for the disc evolution, it is possible to derive the expected surface density profile as a function of the inner disc radius, $R_{\rm in}$, the outer disc radius, $R_{\rm out}$, the radius at which mass is added, $R_{\rm add}$, the rate at which mass is added ${\dot M}_{\rm add}$ and the viscosity $\nu(R)$. For the analytical calculations it is convenient to take the mass input to occur over a region of zero-width, whereas below when presenting numerical simulations we will add mass smoothly over a region with non-zero width. In the former case, \cite{Nixon2021} give the steady solution for $\Sigma(R)$ as
\begin{equation}\label{steady}
\Sigma (R)=\left\{
\begin{aligned}
& \frac{\dot{M}_{\rm add}}{3\pi \nu (R)} \left[ 1 - \left(\frac{R_{\rm in}}{R} \right)^{1/2}\right]\frac{R_{\rm out}^{1/2}-R_{\rm add}^{1/2}}{R_{\rm out}^{1/2}-R_{\rm in}^{1/2}} \ {\rm for} \ R\ \leq \ R_{\rm add} \\
& \frac{\dot{M}_{\rm add}}{3\pi \nu (R)}  \left[ \left(\frac{R_{\rm out}}{R} \right)^{1/2} -1 \right]\frac{R_{\rm add}^{1/2}-R_{\rm in}^{1/2}}{R_{\rm out}^{1/2}-R_{\rm in}^{1/2}} \ {\rm for} \ R\ > \ R_{\rm add}, \\ 
\end{aligned}
\right.
\end{equation}

In this equation we have the viscosity $\nu(R)$, which represents the physical viscosity acting within the accretion disc. In a numerical simulation, as discussed above, we have the total viscosity $\nu_{\rm total} = \nu + \nu_{\rm num}$, where $\nu_{\rm num}$ is the contribution due to the numerical viscosity. We therefore expect that if we start with a disc of surface density given by (\ref{steady}), then as the simulation proceeds the surface density should decay until a new steady state is established corresponding to (\ref{steady}) but with $\nu$ replaced by $\nu_{\rm total}$. By varying the numerical viscosity coefficients, and determining which values lead to the smallest decrease in the expected surface density profile, we can minimise the numerical viscosity. 

\section{Numerical simulations}
\label{sims}
In this section, we provide a description of the code and the setup of our simulations for the circumstellar disc, considering different values of $\alpha_{\rm SPH}$ and $\beta_{\rm SPH}$. Subsequently, we present the results obtained from three different resolutions.

\subsection{Circumstellar disc model}
\label{sta}
\begin{table}
\centering
\begin{tabular}{c c c | c c c c}
\hline
\multicolumn{3}{c|}{\textbf{Model A}} & \multicolumn{4}{c}{\textbf{Model B}} \\
\hline
Model & $\alpha_{\rm SPH}$ & $\beta_{\rm SPH}$ & Model & $\alpha_{\rm SPH}^{\rm min}$ & $\alpha_{\rm SPH}^{\rm max}$ & $\beta_{\rm SPH}$ \\
\hline
A1 & 1.0  & 2.0  & B1  & 0.1  & 1.0 & 2.0 \\
A2 & 0.1  & 2.0  & B2  & 0.01 & 1.0 & 2.0 \\
A3 & 0.01 & 2.0  & B3  & 0.0  & 1.0 & 2.0 \\
A4 & 1.0  & 0.2  & B4  & 0.1  & 1.0 & 0.2 \\
A5 & 0.1  & 0.2  & B5  & 0.01 & 1.0 & 0.2 \\
A6 & 0.01 & 0.2  & B6  & 0.0  & 1.0 & 0.2 \\
A7 & 1.0  & 0.02 & B7  & 0.1  & 1.0 & 0.02 \\
A8 & 0.1  & 0.02 & B8  & 0.01 & 1.0 & 0.02 \\
A9 & 0.01 & 0.02 & B9  & 0.0  & 1.0 & 0.02 \\
   &      &      & BB1 & 0.1  & 1.0 & 2 $\alpha_{\rm SPH}$ \\
   &      &      & BB3 & 0.0  & 1.0 & 2 $\alpha_{\rm SPH}$ \\
   
\hline
\end{tabular}
\caption{Numerical viscosity parameters for model A and model B simulations. For model A simulations both $\alpha_{\rm SPH}$ and $\beta_{\rm SPH}$ are constants. For model B simulations $\beta_{\rm SPH}$ is a constant, and $\alpha_{\rm SPH}$ is allowed to vary between $\alpha_{\rm SPH}^{\rm min}$ and $\alpha_{\rm SPH}^{\rm max}$ following the viscosity switch.}
\label{table:ModelAB}
\end{table}


To understand the effects of different values of $\alpha_{\rm SPH}$ and $\beta_{\rm SPH}$ on simulations of discs, we employ the 3D SPH code \textsc{phantom} \citep{Price2018}. Our simulations involve a single star in the centre with a mass $M$. The accompanying disc, with an initial mass of $0.001 M$, extends from an inner radius of $R_{\rm in}$ = 1 to $R_{\rm out}$ = 10 (we have experimented with an increased outer radius of $R_{\rm out} = 15$ and find our results to be unchanged). The disc is planar and circular. The central star has a spherical accretion radius equal to $R_{\rm in}$ so that any particles which falls within this radius are removed from the simulation. We also impose an outer boundary at $R_{\rm out}$ and remove any particles that move beyond this spherical boundary. These boundary conditions approximate the standard zero-torque boundary conditions for accretion discs \citep[cf.][]{Nixon2021}.\footnote{As the particles arrive at the boundary with a small, but non-zero eccentricity there is a small, but non-zero effective torque at the boundaries. This small effect diminishes with increasing resolution, but can at low-resolution lead to noticeable deviations between the numerical and analytical solutions for discs (see the Appendix of \citealt{Drewes2021} for further discussion).}

We take the disc to be locally-isothermal such that the sound speed of the disc is given by $c_{\rm s}(R) = c_{\rm s,0}(R/R_{\rm in})^{-3/4}$ and the value of $c_{\rm s,0}$ is set such that the disc angular semi-thickness $H/R = c_{\rm s}/v_\phi$ is $0.05$ at the disc inner edge. We model a physical (as opposed to numerical) viscosity in the disc with a Navier-Stokes viscosity corresponding to a Shakura-Sunyaev disc viscosity parameter $\alpha_{\rm SS}=0.1$; this is modelled via the ``two first derivatives'' method outlined in Section 3.2.4 of \cite{Lodato2010}. 

For the radial profile of the disc, we would like to start from the analytical solution for a steady disc given by (\ref{steady}). However, for the SPH simulations we avoid a discontinuous injection function by smoothing the injection of mass over a small radial range. We do this by injecting mass according to a cosine-bell (Hahn function) profile centred at $R_{\rm add}$ and ranging from $R_{\rm add} - \Delta R$ to $R_{\rm add} + \Delta R$, where $\Delta R = 3H(R_{\rm add})$. We take $R_{\rm add}=7$, and ${\dot M}_{\rm add}$ is determined by the required initial disc mass. To determine the (now semi-)analytical solution for $\Sigma(R)$ in this case we numerically integrate the accretion disc diffusion equation for these parameters. The main difference to the analytical solution (\ref{steady}) is a smoothing of the disc surface density at $R=R_{\rm add}$. We employ the semi-analytical solution, shown below as the cyan line in e.g. Fig~\ref{fig:surface}, as the initial conditions for the SPH simulations.

We perform simulations with three different initial particle numbers, $N_{\rm p,\ ini} = 10^5$, $10^6$ and $10^7$, in order to investigate the effects of resolution. Once a steady state is reached the number of particles in the simulations range from $\approx 35,000-65,000$ ($N_{\rm p,\ ini} = 10^5$), $\approx 580,000-780,000$ ($N_{\rm p,\ ini} = 10^6$) and $\approx 7,800,000 - 8,400,000$ ($N_{\rm p,\ ini} = 10^7$). The shell-averaged smoothing length per disc scale-height varies with radius; in the body of the disc from $R\approx 3$ to $R\approx 8$. We find that the shell-averaged smoothing length per disc scale-height $\left<h\right>/H$ varies from $\approx 0.8-1.5$ ($N_{\rm p,\ ini} = 10^5$), $\approx 0.35-0.6$ ($N_{\rm p,\ ini} = 10^6$) and $\approx 0.15-0.4$ ($N_{\rm p,\ ini} = 10^7$).  To explore the effect of changing the viscosity coefficients, we split our simulations into two groups. Models A$N$ (where $N=1,...,9$) employ constant values for both $\alpha_{\rm SPH}$ and $\beta_{\rm SPH}$. Models B$N$ have a constant value for $\beta_{\rm SPH}$ and a variable $\alpha_{\rm SPH}$ with values between $\alpha_{\rm SPH}^{\rm min}$ and $\alpha_{\rm SPH}^{\rm max}$ following the \cite{Cullen2010} viscosity switch as implemented in {\sc phantom} \citep{Price2018}. Table~\ref{table:ModelAB} provides a summary of the specific values of the viscosity coefficients used for each model.


We run the simulation until the disc reaches a steady state, in which the number of active particles fluctuates around a mean value (rather than increasing or decreasing over time). The timescale on which the disc reaches this state is the viscous timescale, given by $t_{\nu} = R^2 / \nu$, where $\nu = \alpha c_{\rm s}H = \alpha (H/R)^2 R^2\Omega$. Putting in the sound-speed profile (with power-law $q=3/4$) and Keplerian rotation velocity we have
\begin{equation}
t_{\nu}(R) = \frac{1}{\alpha} \left(\frac{H_{\rm in}}{R_{\rm in}}\right)^{-2} \left(\frac{R}{R_{\rm in}}\right)^2 \sqrt{\frac{R_{\rm in}^3}{GM}}\,.
\end{equation}
As above, we have $\alpha = 0.1$, $H/R = 0.05$ at the inner edge, an outer edge at $R_{\rm out}=10R_{\rm in}$, and we add mass at a radius $R_{\rm add} =7R_{\rm in}$. The viscous communication timescale from the mass input radius to the inner and outer edges of the disc is similar for these parameters, and is $\sim t_{\nu}(R_{\rm add}) \approx 2.5\times10^5 \sqrt{R_{\rm in}^3/GM}$. We therefore run the simulations for this timescale, which corresponds to $\approx 40,000$ orbits at the disc inner edge, but we note that most simulations settle into a steady state after approximately $30,000$ orbits.

\subsection{Preliminary results}
We observe that SPH simulations of models A$N$ can be separated into four subgroups which have the similar surface density to each other independent of the resolution. These subgroups include (1) the disc with largest $\alpha_{\rm SPH}$ and $\beta_{\rm SPH}$ (model A1), (2) discs with largest $\beta_{\rm SPH}$ (models A2 and A3), (3) discs with the largest $\alpha_{\rm SPH}$ = 1.0 (models A4 and A7) and (4) discs with smaller $\alpha_{\rm SPH}$ = 0.1 and 0.01 (models A5, A6, A8 and A9). The groupings of the simulations arise where the contributions from one or other component of the numerical viscosity is dominating the total numerical viscosity.

Similarly, the B$N$ models can also be separated into subgroups. In this case there are essentially two groups, one in which the $\beta_{\rm SPH}=2$ (the largest $\beta$ value; occurring in models B1, B2 and B3) and the remainder. The smaller number of groups in this case results from employing a viscosity switch on the $\alpha_{\rm SPH}$ term. The shell-averaged value of $\alpha_{\rm SPH}$ is typically of order $0.1$, so setting $\alpha^{\rm min}_{\rm SPH}$ to less than this value has little effect on the results.

Thus, to make the plots in the following section more concise and clearer, we will only show results from models A1, A3, A4 and A9 and models B1, B3, B5 and B9.

\subsection{Simulation results}
\begin{figure*}
    \centering
        \includegraphics[width=8.7cm]{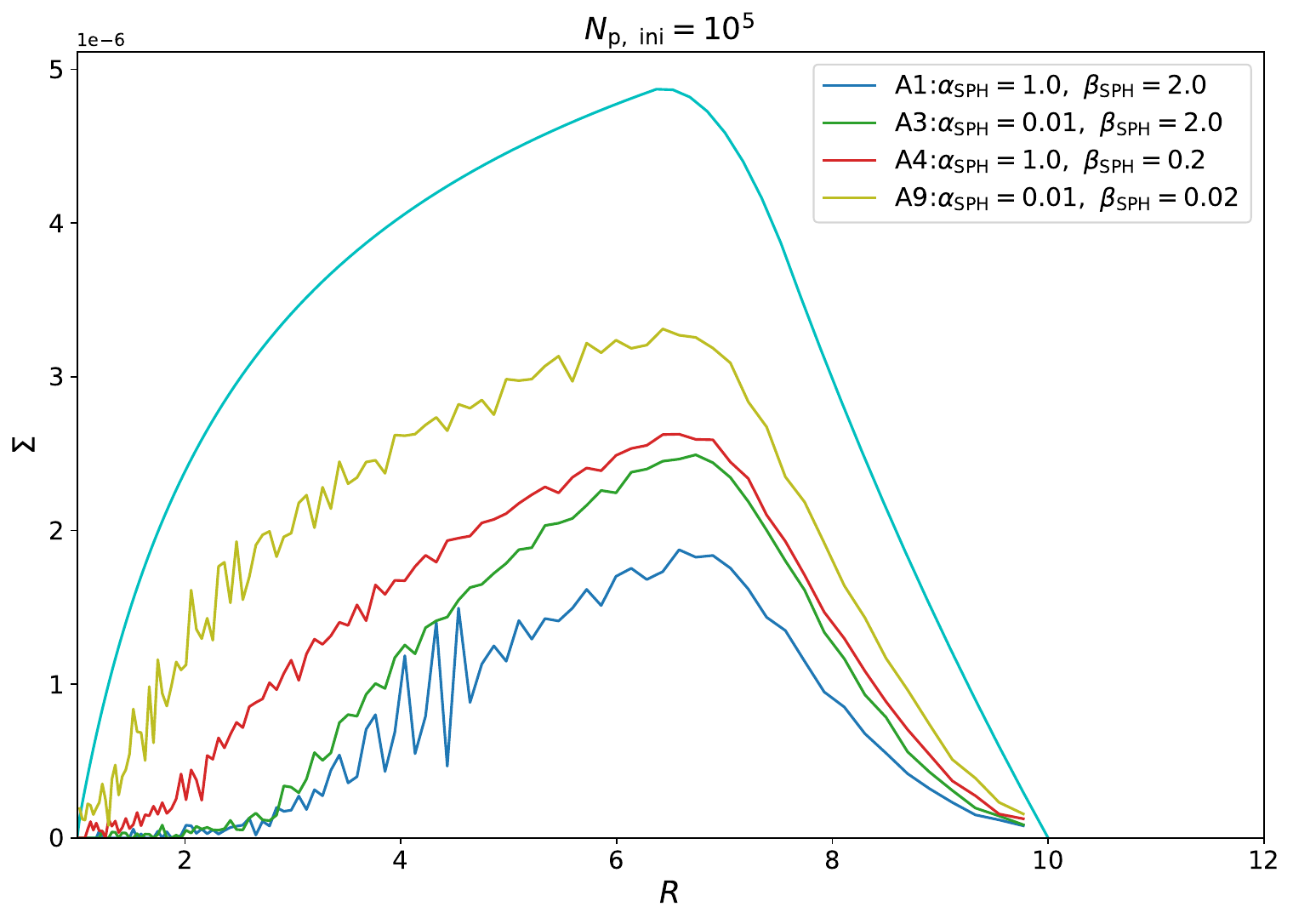}
        \includegraphics[width=8.7cm]{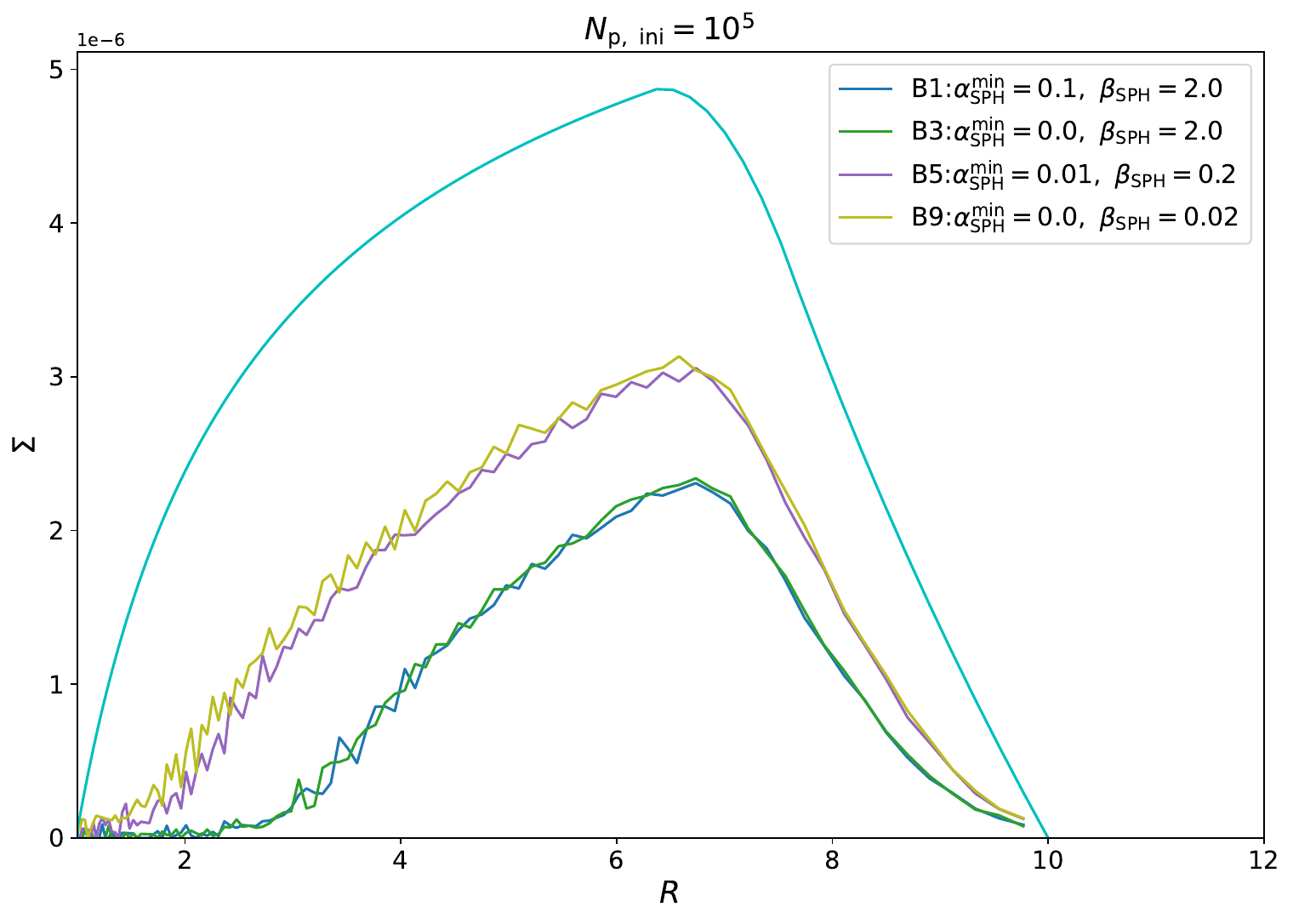}   
        \includegraphics[width=8.7cm]{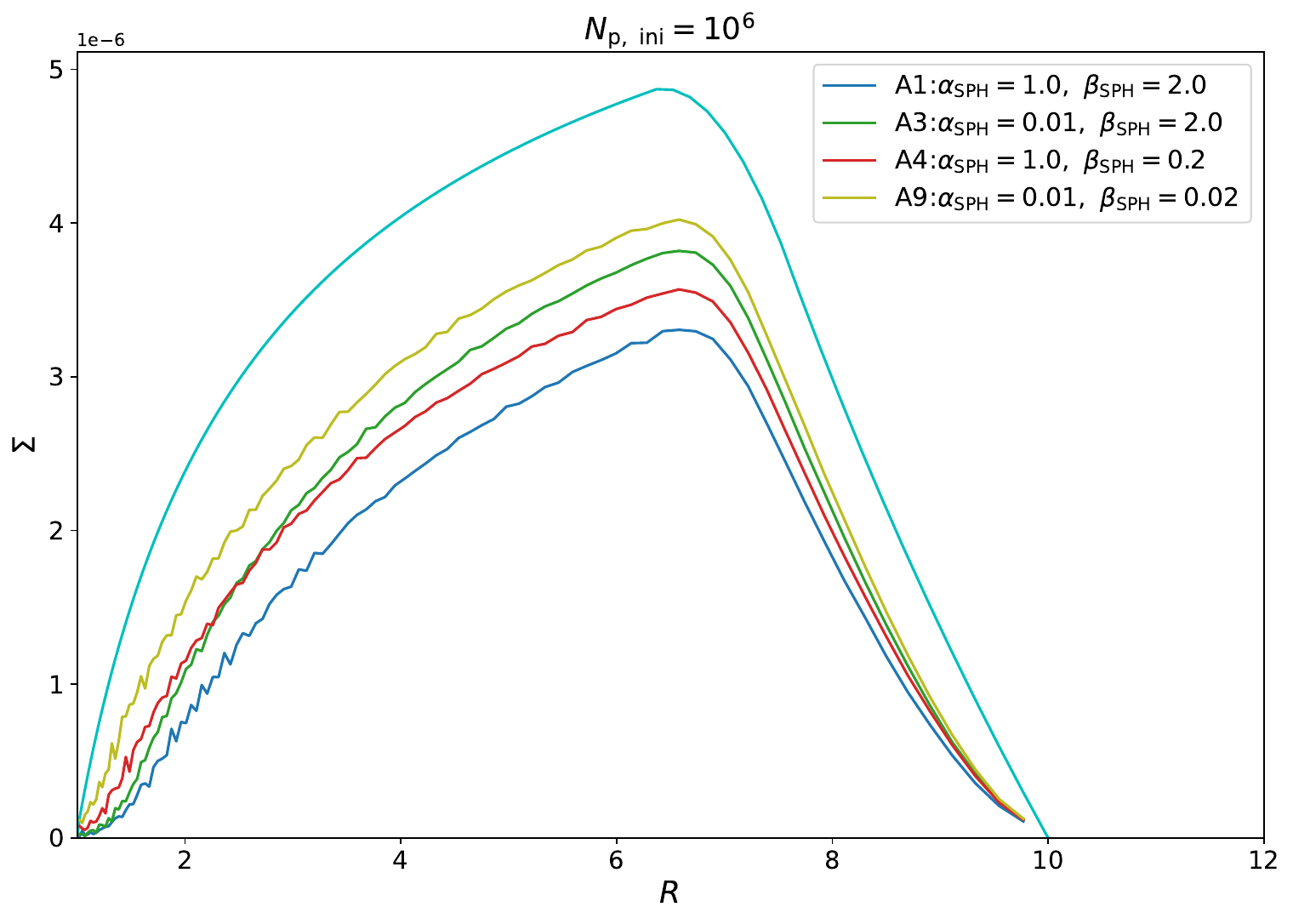}    
        \includegraphics[width=8.7cm]{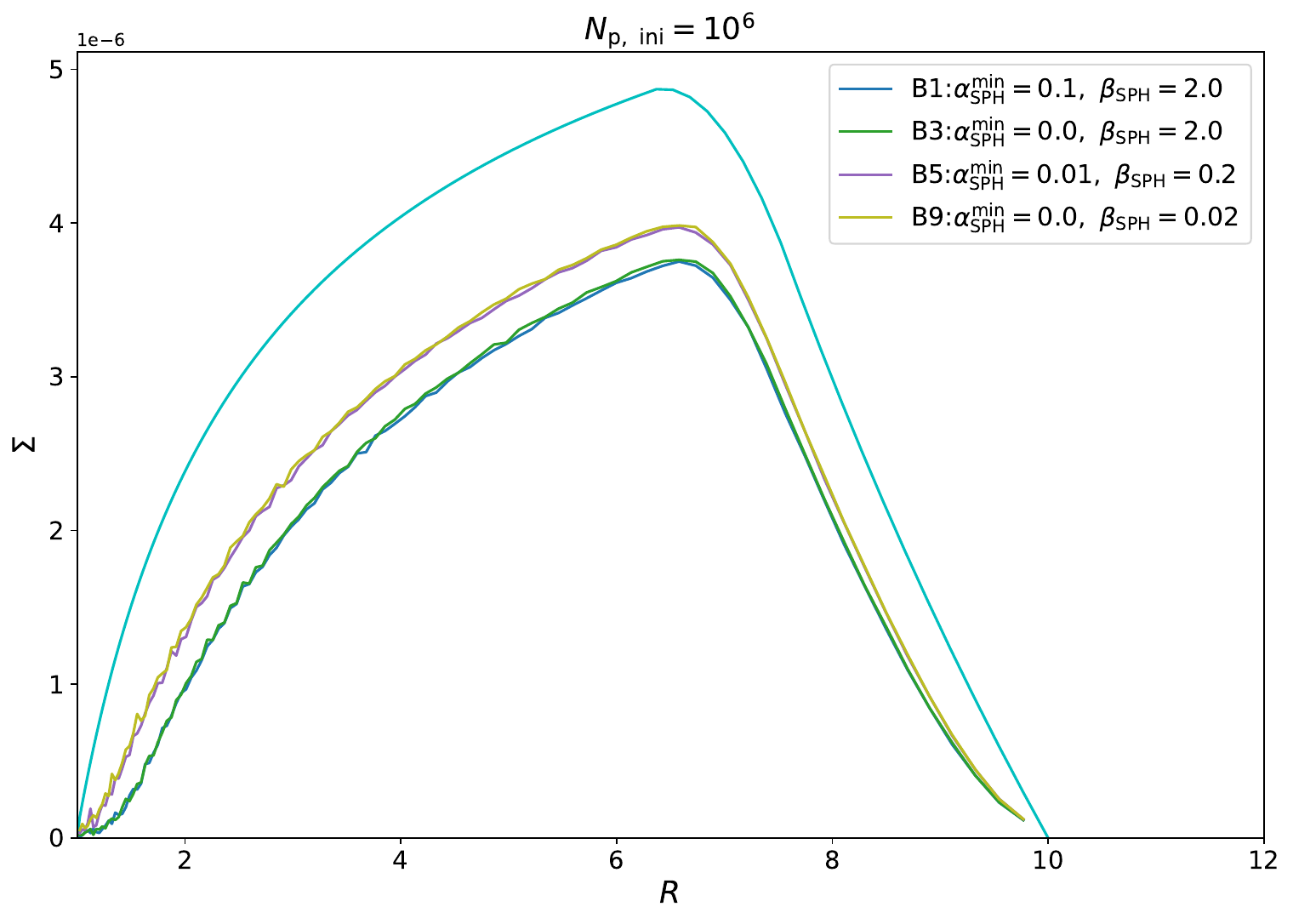}
        \includegraphics[width=8.7cm]{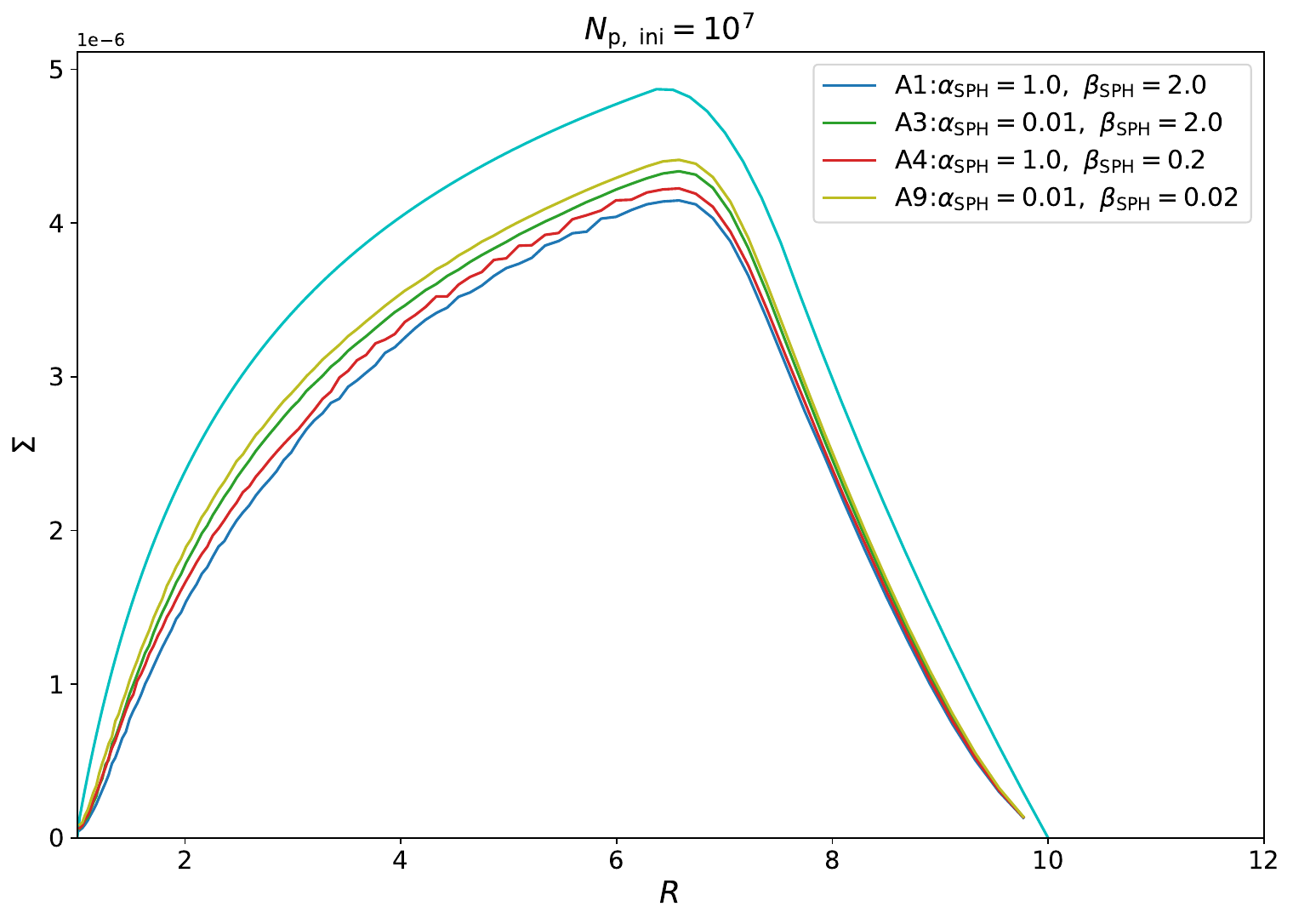}    
        \includegraphics[width=8.7cm]{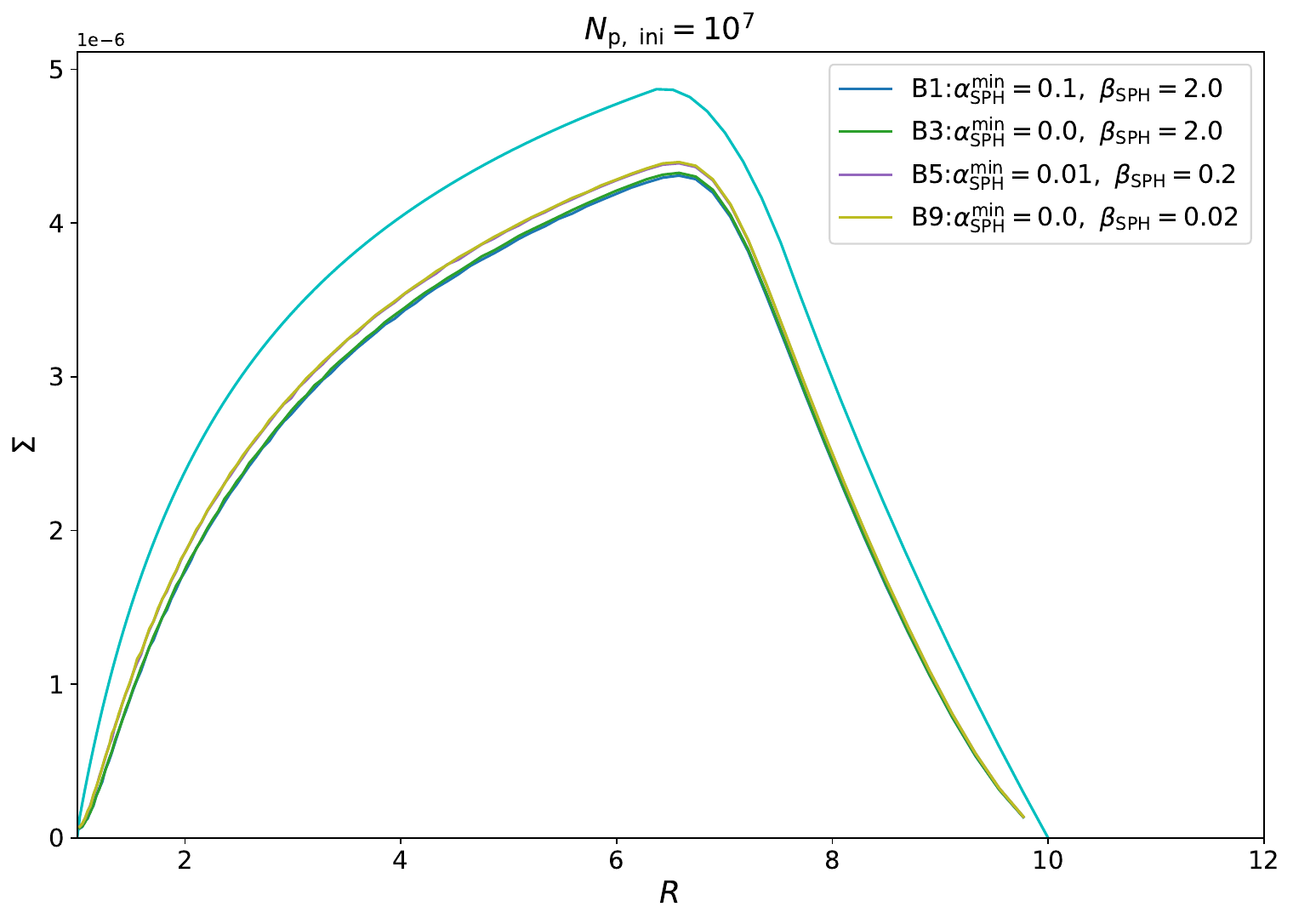}
    \caption{The surface density profiles of each model after they reach a steady state. The left panels contain models A1, A3, A4 and A9  and the right panels contain models B1, B3, B5 and B9. The top panels correspond to the lowest resolution simulations with $N_{\rm p,~ini} = 10^5$. The middle panels show the simulations with $N_{\rm p,~ini} = 10^6$ and the lower panels are  the simulations with $N_{\rm p,~ini} = 10^7$. The cyan line in each panel represents the semi-analytical 1D surface density profile with the same parameters and mass input as the SPH simulations. The steady-state SPH solutions follow a smaller density profile than the semi-analytical solution due to the presence of numerical viscosity. As the resolution is increased, the numerical viscosity decreases and thus its effect diminishes compared to the physical viscosity, and the SPH solutions approach the analytical solution.}
    \label{fig:surface}
\end{figure*}

We plot in Fig.~\ref{fig:surface} the surface density profiles with radius for all selected simulations. To make these plots we bin the particles into 100 radial bins and divide the total mass in each bin by the area of the bin. The left hand panels correspond to models A$N$, and the right hand panels correspond to models B$N$. From top to bottom the resolution increases with $N_{\rm p,~ini} = 10^5$ on the top row, $10^6$ on the middle row, and $10^7$ on the bottom row. 

Starting with the lowest resolution simulations (top panels), we can see that low resolution can lead to a poor match with the analytical solution (the cyan line); both in the shape and the normalisation. The discs with the lowest surface density have $\beta_{\rm SPH}$ = 2.0 for models A1 and A3 but model A3 is higher than A1 since it has lowest $\alpha_{\rm SPH}$ = 0.01.  With lower values of $\beta_{\rm SPH}$ = 0.2 and 0.02, the role of $\alpha_{\rm SPH}$ becomes more prominent. As a result, discs with smaller $\alpha_{\rm SPH} = 0.01$ have the highest surface density (model A9) while discs with the largest $\alpha_{\rm SPH}$ = 1.0 have the smaller surface density (model A4). The shape for model A9 with the lowest levels of dissipation is a reasonable match to the analytical solution (albeit off-set in normalisation). The shape of, for example, models A1 and A3 do not provide a good match to the shape of the analytical solution, particularly in the inner regions. Due to the strong artificial viscosity in the low-resolution cases, the disc mass of the highest surface density case is approximately 61$\%$ of the analytical model (cyan line), while the disc mass with the case with lowest surface density is about 33$\%$ of the analytical model. 

In contrast, model B$N$ in the right hand panel, where the viscosity switch on $\alpha_{\rm SPH}$ is employed, there are only two groups. This is because the viscosity switch reduces the level of dissipation from the linear term and, while $\alpha_{\rm SPH}^{\rm min}$ varies, the shell-averaged $\alpha_{\rm SPH}$ value in these simulations is $\approx 0.1$ . We therefore find two states, one where the quadratic term dominates (when $\beta_{\rm SPH}=2$) and one where the linear term dominates (when $\beta_{\rm SPH} = 0.02-0.2$). 

In the middle-left panel, for the case with $N_{\rm p,\ ini}=10^6$, the role of $\alpha_{\rm SPH}$ becomes more significant in the A$N$ models compared with $N_{\rm p,\ ini}=10^5$. Models A3 exhibit higher surface densities than model A4  even with smaller $\beta_{\rm SPH}$ = 0.2. Model A1, which has the largest $\alpha_{\rm SPH}$ and $\beta_{\rm SPH}$, displays the lowest surface density but it is still higher than that of with the low resolution by a factor of $\approx 1.5$. Model A9 still has the highest surface density of all A$N$ models and is higher than that of the A9 model with lower resolution by a factor of $\approx 1.3$. Overall, high-resolution models have higher surface densities compared to those with lower resolution. For $N_{\rm p,\ ini}=10^6$, the disc mass of the highest surface density is approximately 74$\%$ of the analytical model, while the disc mass of the model with the lowest surface density is around 58$\%$ of the analytical model. Again, in the middle-right panel we only see two groups of simulations, and these are delimited by the value of $\beta_{\rm SPH}$, with those with $\beta_{\rm SPH} = 2$ showing the lower surface density. The difference in the surface density profiles between these two groups is smaller than that for the lower resolution case.

Finally, in the bottom panels we have the simulations with $N_{\rm p,\ ini}=10^7$. These show that each increase in resolution leads to a corresponding increase in the surface density profile and thus a reduction in the numerical viscosity. The difference between model A1 and A9 is very small and the difference between two groups of model B$N$ in the right panel is also very small. Therefore, the effect of choosing specific values of $\alpha_{\rm SPH}$ and $\beta_{\rm SPH}$ become less important once the resolution is high enough for both models A$N$ and B$N$. This occurs once the numerical viscosity is significantly smaller than the physical viscosity we have employed (Shakura-Sunyaev $\alpha = 0.1$). For these high resolution simulations, the disc mass of the highest surface density case (A9, B9) is approximately 84$\%$ of the analytical model while the lowest is about 78$\%$ (A1).

These results show that with increasing resolution, the simulation results more accurately describe the analytical solution. There are different assumptions made by the analytical solution and the numerical simulations (for example, the analytical solution does not account for the influence of pressure gradients which affect, e.g., the angular velocity), so the increase in accuracy at higher resolution implies that these differences are less important than the effect of numerical viscosity at the resolutions employed. The results also demonstrate that increasing resolution leads to decreasing numerical viscosity in the simulations, reinforcing the notion that achieving a sufficiently small numerical viscosity in SPH is possible via application of sufficient resolution (combined with a consistent numerical viscosity scheme). It is also worth noting that the effects of numerical viscosity can also be reduced by application of a larger physical viscosity \citep[cf. Fig.~3 of][]{Lodato2010}.

\section{Discussion} 
\label{discussion}

The simulations presented above show a variety of features that we discuss here. First at low resolution the details of the numerical viscosity (the values of the linear and quadratic parameters and whether or not the linear term employs a viscosity switch) can play an important role in determining the disc surface density profile (see e.g. top panels of Fig. 1). However, as the resolution is increased, the level of numerical viscosity in all cases is reduced such that the dominant viscosity is that imposed by the physical (Navier-Stokes) viscosity term (e.g. lower panels of Fig. 1). However, there is clearly still some benefit in finding the optimal values for the viscosity coefficients. 

Typically we have found that $\beta \approx 0.2$ results in better surface density profiles than the default value of $\beta = 2$. However, in the simulations we have performed the disc is smooth and planar. In many discs, and other astrophysical flows, there are more complex structures and, in particular, shocks that can require $\beta \approx 2$ to properly model the flow \citep[e.g.][]{Meru2012}. This issue has been remedied for the linear term by the application of a switch which reduces the value of the coefficient in regions of the flow where shocks are not detected \citep[e.g.][]{Morris1997,Cullen2010}. In {\sc phantom} the switch is only applied to the linear term and not the quadratic term. We return to this point below.

It is widely accepted that $\left<h\right>/H$, must be smaller than unity to properly model an accretion disc. This is in part due to the effective viscosity arising from the numerical viscosity terms, which are linearly and quadratically proportional to this quantity. But primarily this is due to the need to resolve gradients in the pressure, and in an accretion disc the pressure length-scale is $H$. As we have seen above, simulations with $\left<h\right>/H \lesssim 1$ can recover an acceptable shape for the surface density profile of a disc (albeit with a large numerical viscosity), but in many cases the local dynamics in the disc can require $\left<h\right>/H \ll 1$ \citep[see, e.g.,][for an example]{Drewes2021}.

The ratio of the effective viscosities arising from the numerical terms, using the formulae provided by \cite{Meru2012}, is
\begin{equation}\label{eq:numratio}
    \frac{\alpha_{\rm num}^{\rm quad}}{\alpha_{\rm num}^{\rm lin}} = \frac{135}{62\pi}\frac{\beta_{\rm SPH}}{\alpha_{\rm SPH}}\frac{\left<h\right>}{H}\,.
\end{equation}

From this equation, it is straightforward to see that when $\beta_{\rm SPH} \approx \alpha_{\rm SPH}$ and the disc is well-resolved with $\left<h\right>/H < 1$, then the contribution from the quadratic viscosity is smaller than the contribution from the linear viscosity. For example, with the classical $\alpha_{\rm SPH} = 1$ and $\beta_{\rm SPH} = 2$, the resolution required to make this ratio less than unity is $\left<h\right>/H < 62\pi/270 \approx 0.72$. Consequently\footnote{ An additional reason is the prevalence of employing the ``artificial viscosity for a disc'' method \citep[Section 3.2.3 of][]{Lodato2010} which uses the {\it initial conditions} of the disc to scale the linear numerical viscosity term to the magnitude of the desired Shakura-Sunyaev viscosity. We discuss this method in the last paragraph of Section~\ref{discussion}.} it is common to present only the numerical contribution to the total viscosity arising from the linear term, with the quadratic term assumed to be small. Further, for this reason, the standard approach (e.g. employed by {\sc phantom}) when employing a viscosity switch (e.g. Morris-Monaghan or Cullen-Dehnen) is to only apply the switch to $\alpha_{\rm SPH}$, leaving $\beta_{\rm SPH}$ a constant.

However, when employing a viscosity switch that reduces the value of $\alpha_{\rm SPH}$ in regions of the flow away from shocks it is possible---and for well-resolved flows typical---for the shell-averaged value of $\alpha_{\rm SPH}$ to fall to values $\approx 0.1$. In this case, from (\ref{eq:numratio}), the resolution requirement to ensure the quadratic term contributes less than the linear term is $\left<h\right>/H \lesssim 0.1$. This is a sufficiently stringent condition that very few simulations reported in the literature meet this requirement. Similarly, when employing the ``artificial viscosity for a disc'' method of \cite{Lodato2010}, it is common for the scaled (constant) value of $\alpha_{\rm SPH}$ to be of order $0.1$ (with $\beta_{\rm SPH}=2$ applied by default).

It is worth noting that both of the switches presented by \cite{Morris1997} and \cite{Cullen2010} are presented with both the linear and quadratic terms employing the switch. In this instance the requirement for the linear term to contribute more than the quadratic one returns to $\left<h\right>/H \lesssim 0.72$ (assuming the ratio of $\beta_{\rm SPH}/\alpha_{\rm SPH} = 2$). As this requirement effectively amounts to ensuring that the pressure length-scale is resolved, most simulations should satisfy this. It is therefore worth seeing if simply multiplying the quadratic term by the switched $\alpha_{\rm SPH}$ value yields a better outcome. As we have seen above, when including the switch on $\alpha_{\rm SPH}$ the results are essentially independent of our choice of $\alpha_{\rm min}$; this is because we have taken values of 0, 0.01 and 0.1, and with the viscosity switch active the shell-averaged value of $\alpha_{\rm SPH}$ is typically $\gtrsim 0.1$.\footnote{Note that the shell-averaged value of $\alpha_{\rm SPH}$ decreases with increasing resolution; see also \cite{Drewes2021}.} 

Including a switch on the $\beta_{\rm SPH}$ term, effected by making $\beta_{\rm SPH}$ a multiple of (the switched) $\alpha_{\rm SPH}$, may serve to minimise the numerical viscosity, while allowing the numerical viscosity to grow to the required values in simulations that encounter strong shocks.\footnote{While \cite{Price2010} find that, with a switch active on both linear and quadratic terms, a value of $\beta_{\rm SPH} = 2$ was not sufficient to resolve strong shocks, they employed the Morris-Monaghan switch. \cite{Cullen2010} note that their switch, also active on both the linear and quadratic terms, can model strong shocks appropriately with $\beta_{\rm SPH} = 2$. It would therefore be worthwhile repeating the simulations of \cite{Price2010} with the Cullen-Dehnen switch to see if the requirement of a larger $\beta_{\rm SPH}$ value is necessary.} 

\begin{figure}
    \centering
    \includegraphics[width=8.4cm]{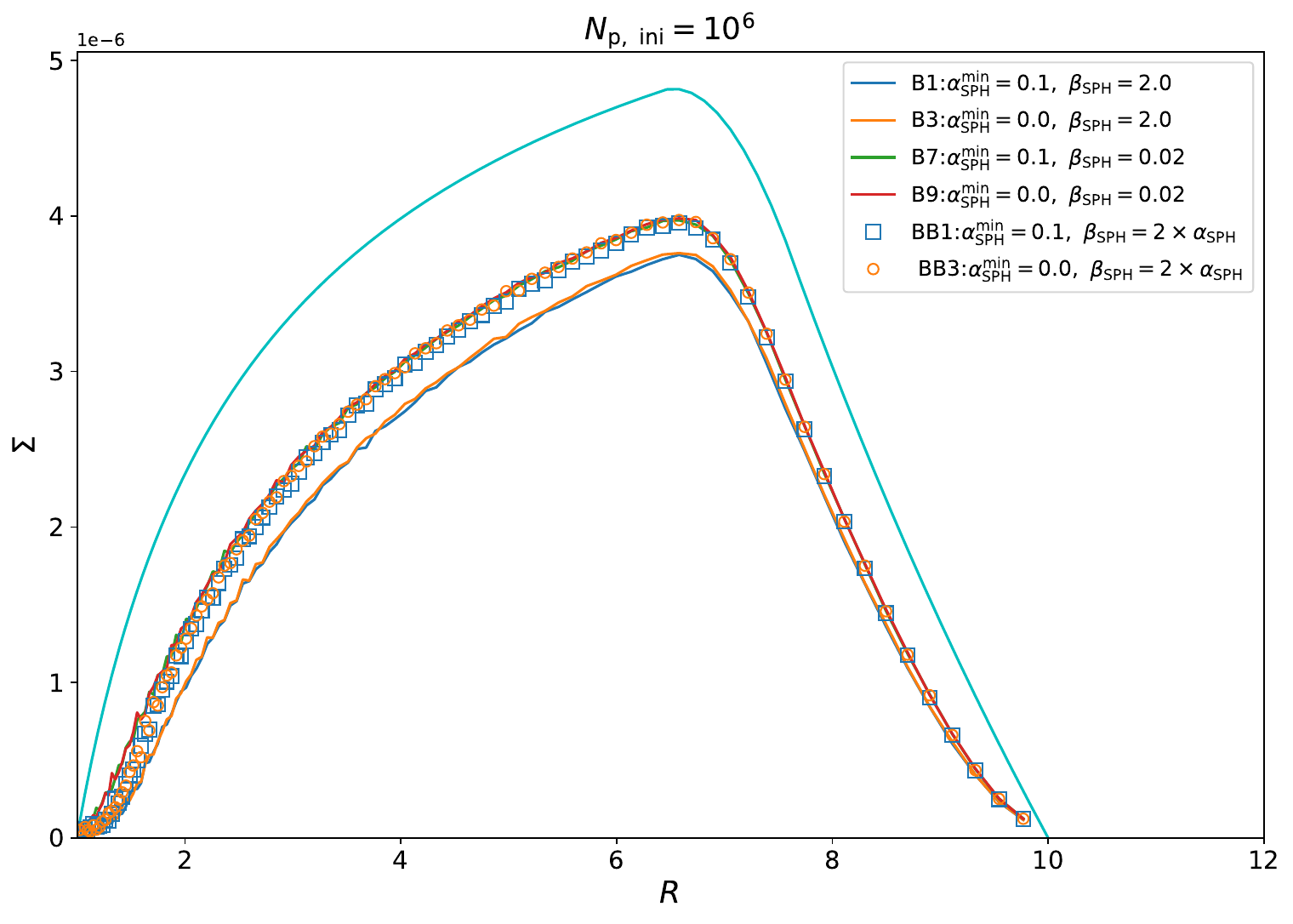}
    \caption{Dotted lines are surface density profiles of models B1(blue), B3(yellow), B7(green) and B9(red) with $N_{\rm p,~ini}=10^6$. Blue squares and yellow circles are similar to models B1 and B3 except $\beta_{\rm SPH}=2\times \alpha_{\rm SPH}$.}
    \label{fig:BB}
\end{figure}

As an example, we provide in Fig.~\ref{fig:BB} examples of the simulations reported above but this time with $\beta_{\rm SPH}=2\alpha_{\rm SPH}$ (models BB1 and BB3). These simulations show surface density profiles that match the highest profiles we have found by varying the numerical viscosity coefficients.\footnote{ Within $R$ < 2.0, there are minor differences where the surface densities of models B7 and B9 are slightly higher than those of models BB1 and BB3. This occurs because the innermost regions, where there are relatively few particles, have an average value of $\alpha_{\rm SPH}$ such that $2\alpha_{\rm SPH}$ is larger than 0.02.} This is significant because this approach -- taking $\beta_{\rm SPH}$ to be a multiple of the ``switched'' $\alpha_{\rm SPH}$ -- allows the numerical viscosity to be minimised in relatively simple flows that do not require strong numerical viscosity while also allowing the numerical viscosity to grow to required values in the presence of strong shocks. This approach also reduces the number of parameters that need to be set by the user. As suggested by \cite{Cullen2010} and evidenced by Fig.~\ref{fig:BB}, we can take $\alpha_{\rm SPH}^{\rm min} = 0$, and then we only need to choose a maximum value for $\alpha_{\rm SPH}$ and the ratio between the two terms. Here we have taken $\alpha_{\rm SPH}^{\rm max} = 1$ and the ratio $\beta_{\rm SPH}/\alpha_{\rm SPH} = 2$. These values are likely to be good choices for a wide range of numerical simulations, but there are also likely to be some cases where these values could be further improved. 

Additionally, to check that our suggested approach is an improvement upon the widely used ``artificial viscosity method for a disc'' of \cite{Lodato2010} we include an example simulation using the same setup as here but with this method. The results are presented in Appendix~\ref{app}. These additional simulations show that the ``artificial viscosity method for a disc'' does not correctly capture the shape of the disc surface density profile. There are additional reasons for avoiding this methodology for modelling accretion disc viscosity. These include: (i) any subsequent evolution of the disc surface density changes the local resolution, and thus the numerical viscosity, rendering the scaling incorrect, (ii) unless a specific combination of power-law surface density and sound speed profiles are employed the effective viscosity parameter, $\alpha$, becomes a function of radius rather than the desired constant \citep[see e.g. equations 22 \& 23 of][]{Lodato2007b}, (iii) requesting a low value of the physical viscosity can scale the numerical viscosity to too small a value to provide an accurate numerical solution, (iv) the contribution from the quadratic viscosity has not been included in the scaling of this method in the repository version of {\sc phantom} to date meaning that the effective viscosity is larger than reported, and (v) the numerical viscosity terms generate both a shear and bulk viscosity with the magnitude of the bulk viscosity {\it larger} than the shear viscosity \citep[e.g. equations 36 \& 37 of][]{Lodato2010}, thus when attempting to model a shear viscosity it is best to do so with an explicit shear viscosity and minimised numerical viscosity to avoid an excessive and unphysical bulk viscosity.

\section{Conclusions}
\label{conclusion}
In this paper, we have simulated circumstellar discs with different values of the artificial viscosity coefficients, $\alpha_{\rm SPH}$ and $\beta_{\rm SPH}$, in the SPH code {\sc phantom} with the goal of minimising the numerical viscosity present in simulations of discs. Naively one might set the coefficients to small values to achieve this, but this results in poor ordering of the particles and thus additional numerical dissipation \citep[e.g.][]{Meru2012}. Setting the coefficients too high also results in excessive dissipation, and as such an optimal value for planar, smooth discs occurs at around $\alpha_{\rm SPH}= 0.1$ and $\beta_{\rm SPH}=0.2$.\footnote{Note that in Fig.~\ref{fig:surface} this case is not plotted, but it overlaps essentially exactly with the A9 model where $\alpha_{\rm SPH}= 0.01$ and $\beta_{\rm SPH}=0.02$} This is consistent with, for example, \cite{Lodato2004,Meru2012}. However, in more general cases, larger values of the coefficients are required and thus, for example, the default values in {\sc phantom} are (maximum) $\alpha_{\rm SPH} = 1.0$ and (constant) $\beta_{\rm SPH}=2.0$. We therefore advocate, in agreement with \cite{Morris1997} and \cite{Cullen2010}, that the viscosity switches that are designed to reduce the numerical viscosity in regions away from shocks be applied to both the linear and quadratic numerical viscosity coefficients in future simulations. This allows the numerical viscosity to act in regions where it is required, but not over exert itself in regions where it is not. We also advocate that SPH simulations of discs that include a physical Shakura-Sunyaev viscosity do so with an explicit Navier-Stokes viscosity (see Section 3.2.4 of \citealt{Lodato2010}), as opposed to the ``artificial viscosity for a disc'' method (Section 3.2.3 of \citealt{Lodato2010}). Whether or not the numerical viscosity is then playing a significant role in the dynamics of the simulation can be checked through, e.g., the use of resolution studies.

\section*{Data Availability}
The simulations presented in this paper made use of the {\sc phantom} SPH code (Astrophysics Source Code Library identifier ascl.net/1709.002 and the code repository \url{https://github.com/danieljprice/phantom}). The data underlying this article will be shared on reasonable request to the corresponding author.

\section*{Acknowledgements}
We thank Daniel Price for useful discussions. CC and CJN acknowledge support from the Science and Technology Facilities Council (grant number ST/Y000544/1). CJN acknowledges support from the Leverhulme Trust (grant number RPG-2021-380). This project has received funding from the European Union’s Horizon 2020 research and innovation program under the Marie Skłodowska-Curie grant agreement No 823823 (Dustbusters RISE project). This work was performed using ARC4, part of the High Performance Computing facilities at the University of Leeds, UK and the DiRAC Data Intensive service at Leicester, operated by the University of Leicester IT Services, which forms part of the STFC DiRAC HPC Facility (www.dirac.ac.uk). The equipment was funded by BEIS capital funding via STFC capital grants ST/K000373/1 and ST/R002363/1 and STFC DiRAC Operations grant ST/R001014/1. DiRAC is part of the National e-Infrastructure.



\bibliographystyle{mnras}
\bibliography{main} 

\appendix
\section{Disc viscosity using the artificial viscosity term} 
\label{app}
\cite{Lodato2010} suggest including a Shakura-Sunyaev disc viscosity either by a direct Navier-Stokes viscosity or by adapting the numerical viscosity term to mimic a Shakura-Sunyaev viscosity. We have described above (in footnote 7) why the latter might not be the best approach to modelling disc viscosity in SPH simulations. Here, we provide a comparison between the two methods for the steady-state discs we simulate here.

\begin{figure*}
    \centering
    \includegraphics[width=8.3cm]{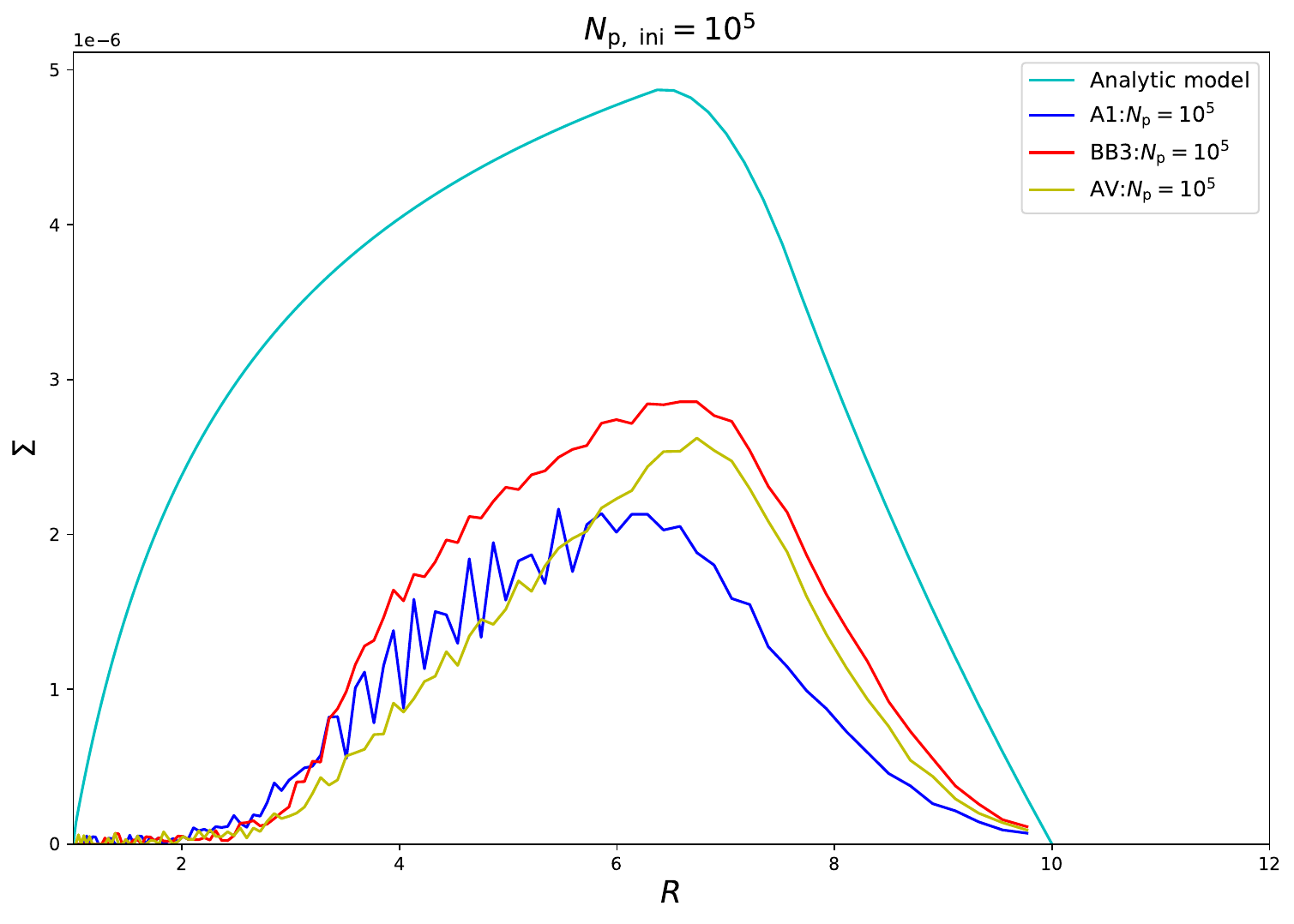}
    \includegraphics[width=8.3cm]{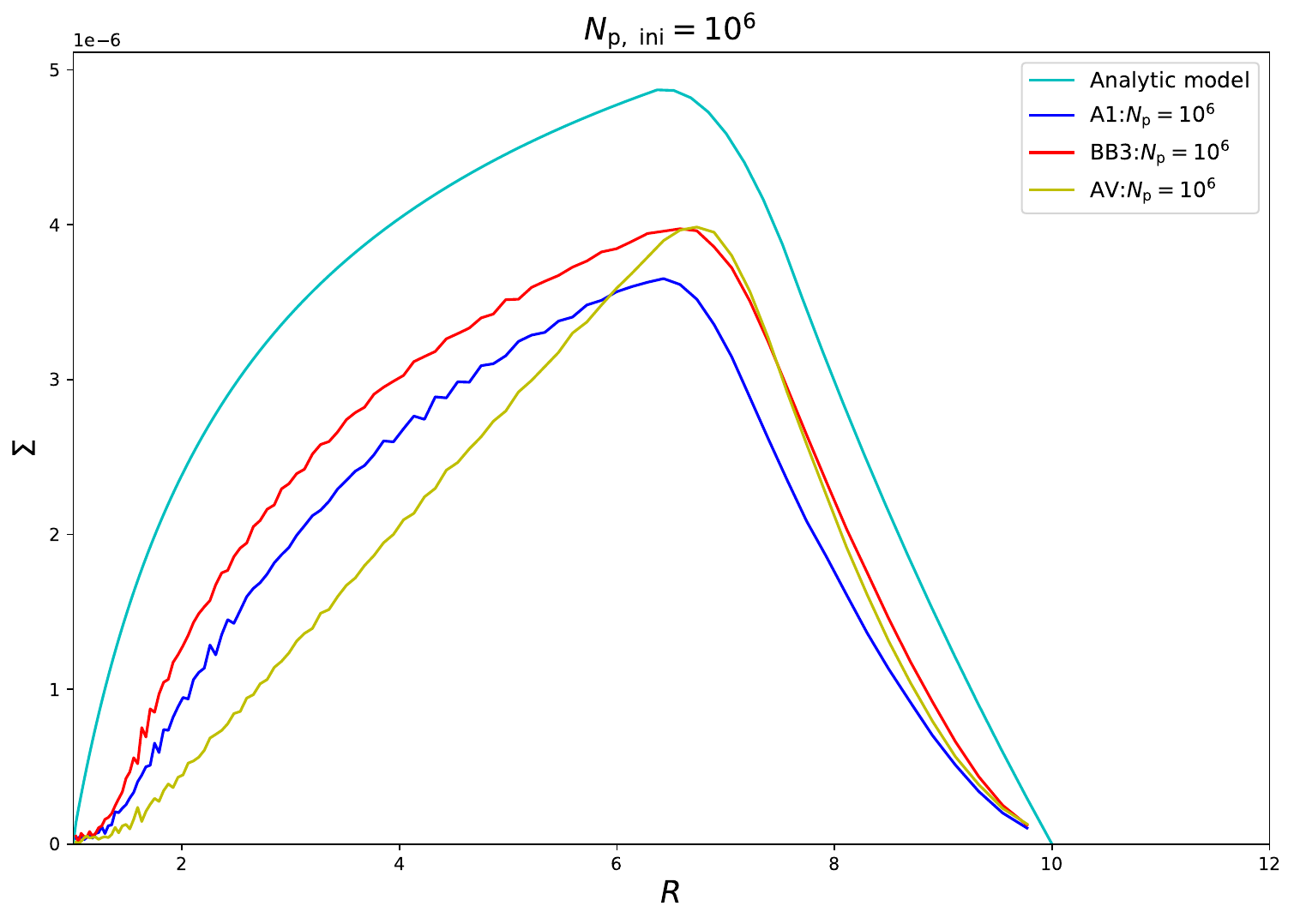}
    \cprotect\caption{Comparison of the Navier-Stokes physical viscosity method employed in the main text and the ``artificial viscosity for a disc'' method (e.g. Section 3.2.3 of \citealt{Lodato2010}) that is widely employed in {\sc phantom} simulations of discs. Each plot shows the surface density profile of the analytic model (cyan), two example simulations from the main text, (A1 given by the blue line; for which $\alpha_{\rm SPH}=1$ and $\beta_{\rm SPH}=2$ are both constants, and BB3 given by the red line; for which $\alpha_{\rm SPH}^{\rm min}=0$, $\alpha_{\rm SPH}^{\rm max}=1$ and $\beta_{\rm SPH}=2\alpha_{\rm SPH}$), and a simulation using the artificial viscosity terms to model the physical viscosity (enabled in the {\sc phantom} code with the option \verb|DISC_VISCOSITY=yes|) denoted AV (yellow lines). The left panel has $N_{\rm p, ini} = 10^5$ while the right panel has $N_{\rm p, ini} = 10^6$.}
    \label{fig:AV}
\end{figure*}

We performed two additional simulations with the same initial disc conditions but using the ``artificial viscosity for a disc'' method (Section 3.2.3 of \citealt{Lodato2010}) to model the disc viscosity. In these tests we adopt the {\sc phantom} default value of $\beta_{\rm SPH} = 2$, and we calculate the $\alpha_{\rm SPH}$ value using the same method as employed in the disc setup in the default version of {\sc phantom} \citep[i.e. applying equation 38 of][to the initial conditions]{Lodato2010}. This results in values of $\alpha_{\rm SPH} = 1.54$ and $\alpha_{\rm SPH}=3.33$ for the $N_{\rm p,ini}=10^5$ and $N_{\rm p,ini}=10^6$ simulations respectively. In Fig.~\ref{fig:AV} we plot the results alongside the results of models A1 and BB3 from the main text, with the left hand panel showing the case with $N_{\rm p,ini} = 10^5$ and the right hand panel showing the case with $N_{\rm p,ini} = 10^6$. We note that model A1 provided the poorest fit to the analytical solutions out of all of the simulations we present in the main text, while BB3 represents the method we propose based on our findings in the main text. Fig.~\ref{fig:AV} demonstrates that, while in each case (both resolutions) the ``artificial viscosity for a disc'' method achieves a peak surface density (near the location at which particles are injected) that is similar to the peak in the BB3 simulation, the AV method generates solutions that do not conform to the expected shape of the analytical solution. Further, this does not improve as the resolution is increased.

Finally, we have also tried using $\alpha_{\rm SPH} = 0.0$ and $\beta_{\rm SPH} = 10$ as this approach could negate the need for switches as suggested by \citet[][we thank D. Price for suggesting this]{Price2024}, but we find that this does not provide a good approach to modelling discs at the resolutions we have employed.

\bsp	
\label{lastpage}
\end{document}